\documentclass[prl,aps,amssymb,twocolumn,superscriptaddress,showkeys,notitlepage,nobalancelastpage]{revtex4-1}
\usepackage{graphicx}
\usepackage{dcolumn}
\usepackage{bm}
\usepackage{ulem}
\usepackage[caption=false]{subfig} 
\usepackage{changepage}
\usepackage{float}
\usepackage{amsmath}
\usepackage{xcolor}
\usepackage{bbold}
\usepackage{mathtools}

\usepackage{dsfont}

\usepackage{hyperref}
\hypersetup{
    colorlinks=true,
    linkcolor=red,
    urlcolor=black,
    citecolor=black
}

\newcommand{\FGT}{Fe$_3$GeTe$_2$ }
\newcommand{\kvec}{\mathbf{k}}

\newcommand{\ket}[1]{| #1\rangle}
\newcommand{\bra}[1]{\langle #1 |}

\begin{document}

\begin{abstract}
\FGT has gained attention in the condensed matter community for its potential to be exfoliated into thin films with ferromagnetic (FM) order, thanks to its van der Waals layered structure and significant intrinsic anomalous Hall conductivity (AHC). In this work, we analyze the electronic structure and show that, contrary to prior claims, the bulk of the AHC cannot arise from gapped nodal lines. By studying the material’s symmetry properties, both with and without spin-orbit coupling (SOC) and across paramagnetic and FM phases, we find that \FGT hosts mirror-symmetry-protected nodal lines, which support surface drumhead states. Additionally, we identify three key sources of AHC: nodal lines in the paramagnetic phase gapped by the FM order, Weyl points within specific energy ranges, and gaps between spin-up and spin-down bands caused by SOC. Finally, our calculations suggest that electron doping could increase the AHC up to four times compared to its value at the computed Fermi level.
\end{abstract}

\title{Origins of the anomalous Hall conductivity in the symmetry enforced \FGT nodal-line ferromagnet.}

\author{Mikel García-Díez}
\affiliation{Donostia International Physics Center, Paseo Manuel de Lardizabal 4, 20018 Donostia-San Sebastian, Spain.}
\affiliation{Physics Department, University of the Basque Country (UPV/EHU), Bilbao, Spain}
\author{Haim Beidenkopf}
\affiliation{Department of Condensed Matter Physics, Weizmann Institute of Science, Rehovot, Israel.}
\author{I\~nigo Robredo}
\affiliation{Luxembourg Institute of Science and Technology (LIST), Avenue des Hauts-Fourneaux 5, L-4362 Esch/Alzette, Luxembourg}
\author{M. G. Vergniory}
\affiliation{Donostia International Physics Center, Paseo Manuel de Lardizabal 4, 20018 Donostia-San Sebastian, Spain.}
\affiliation{
Département de Physique et Institut Quantique,
Université de Sherbrooke, Sherbrooke, J1K 2R1 Québec, Canada.
}

\date{\today}

\maketitle

\section{Introduction}

\FGT is an itinerant ferromagnet with a Curie temperature $T_c\approx 220K$ and a hexagonal layered structure\cite{structure}. The bulk material is formed by slabs of Fe$_3$Ge lying between layers of Te bound by weak van der Waals forces to form a 3D structure, which makes it easily exfoliable\cite{exfoliate} or grown by deposition\cite{beam_epitaxy}. The two non-equivalent Fe positions in the unit cell display a local magnetic moment of around $1.4\mu_B$ per atom along the stacking $c$ axis\cite{magnetic_moment, magnetic_properties}. Previous studies have shown that \FGT displays a large  anomalous Hall conductivity\cite{experimental_ahc}, anomalous Nernst effect\cite{nernst_effect} and non-linear quantum Hall effect\cite{second_harmonic}, which has drawn the attention for their applicability in the development of novel spintronic devices. Its magnetic properties have been studied in-depth\cite{magnetic_properties1,magnetic_properties2,magnetic_properties3,magnetic_properties4,magnetic_properties5}, discovering the possibility to tune its magnetic configuration via strain\cite{tunable_strain} or applied voltage\cite{tunable_voltage}. Moreover, other studies claim that \FGT shows Kondo lattice behavior\cite{kondo_lattice} and skyrmionic phases\cite{skyrmion}.

Full knowledge of the symmetry of \FGT is necessary to understand its topological properties. Indeed, while topological phases may arise even in the absence of crystalline symmetries, the latter offers many mechanisms to protect the stability of topological features such as band crossings giving rise to Dirac and Weyl nodes or nodal-line semimetals\cite{protected_weyls,composite_weyls,dirac_protected}. \FGT offers another platform to investigate the interplay between topology and symmetry in magnetically ordered materials, since its ground-state is described by a magnetic space group (MSG) or a spin space group\cite{zhida_ssg} (SSG) when spin orbit coupling (SOC) has a negligible effect. The non-zero local magnetic moments break time reversal symmetry (TRS) intrinsically, a necessary condition for some specific features to appear. For example, centrosymmetric crystals cannot host Weyl nodes when TRS is present\cite{weyl_requirements}.

Previous research also shows that \FGT hosts one nodal line located very near the Fermi level, when SOC is neglected\cite{kh_nodal_line}. In particular, the band structure shows a twofold-degenerate crossing at the $K=(1/3,1/3,0)$ point. This degeneracy extends along the high-symmetry $P$ line joining the $K$ and $H=(1/3,1/3,1/2)$ wave vectors. When SOC is taken into account, the nodal line was shown to be gapped, leading to an avoided crossing with a very small energy gap along which Berry curvature is concentrated, which results in a large flux. This, in turn, induces a large anomalous Hall conductivity (AHC), which was forbidden by the TRS in the non-magnetic structure. This mechanism was put forward to explain the high anomalous Hall conductivity reported in experiments. 

In this article, we perform an exhaustive symmetry analysis of the \FGT electronic band structure with and without SOC. In this way, we are capable of describing the aforementioned lifting of the degeneracy, but also of uncovering more topological features in this compound.
Indeed, building upon previous research, we find that \FGT still hosts \textit{many} symmetry-protected nodal lines, some of which are near the Fermi energy, even when SOC is taken into account. We also identify clear drum-head states from a nodal line close to the Fermi level. Guided by the symmetry analysis,  we are able to establish the main contributions to the AHC. Computing the AHC for different values of the chemical potential reveals that three mechanisms are required to explain the full response: nodal lines gapped by the ferromagnetic order and SOC, Weyl nodes and avoided crossings due to SOC.

This article is structured as follows. In Section \ref{sec:symmetry}, we give a full analysis of the symmetry of the electronic band structure without SOC (Subsec.\ref{ssec:sym_nosoc}) and with SOC (Subsec.\ref{ssec:sym_soc}). In Section \ref{sec:nodal_lines} we explain the symmetry protection of the nodal lines (Subsec.\ref{ssec:nodal_abundance}), giving some examples around the Fermi energy and localizing drum-head states (Subsec.\ref{ssec:drumhead}). Section \ref{sec:ahc} explains the main three contributions to the AHC, which are due to gapped nodal lines (Subsec.\ref{ssec:broken_nodal}), Weyl nodes (Subsec.\ref{ssec:weyl_points}) and SOC-induced gaps (Subsec.\ref{ssec:soc_gaps}). Finally, Section \ref{sec:conclusion} summarizes the main results of this work.

\section{Methods}

In order to study the electronic band structure of \FGT we performed Density Functional Theory (DFT) calculations
as implemented in the \textit{Vienna Ab-initio Simulation Package} (VASP) software\cite{vasp1,vasp2}. We used Projector Augmented Wave (PAW) pseudo-potentials\cite{paw} with the Perdew-Burke-Ernzenhof (PBE) implementation of the Generalized Gradient Approximation (GGA) for the exchange-correlation functional\cite{gga_pbe}. Additionally, van der Waals forces were included using the DFT-D3 method with Becke-Johsnon damping function\cite{vdw}. The energy cut-off of the plane wave basis was set to 600 eV
and the $\kvec$ point grid was a Gamma-centered, regular Monkhorst-Pack point set of dimensions $11\times 11\times 7$ for a uniform density of points given the unit-cell taken from the reference literature\cite{structure}. The symmetry identification and analysis was performed using the IrRep package\cite{irrep}. Additional information about the space group symmetry was retrieved from the Bilbao Crystallographic Server\cite{double_space_group,mtqc}. Consequently, we will adopt the same labeling conventions and choice of standard unit cell. We constructed maximally localized Wannier functions as implemented in Wannier90 \cite{wannier90}, with a starting basis of $d$ orbitals for Fe and $p$ for Te and Ge, with a frozen energy window of 3 eV centered at the Fermi level.
Bulk and surface state calculations were performed using the interpolated Wannier TB model as implemented in WannierTools\cite{wanniertools} and AHC calculations were run using WannierBerri\cite{wannierberri}.
\section{Symmetry analysis of the band structure} \label{sec:symmetry}
\begin{table}[h!]\label{tab:crystal_str}
	\begin{center}
	\begin{tabular}{cccc}
		\hline
		\multicolumn{4}{c}{Lattice vectors}                   \\ \hline
		Vector    & $\mathbf a_1$        & $\mathbf a_2$        & $\mathbf a_3$       \\
		Length (Ang.)    & $3.991$ & $3.991$    & $16.33    $ \\ \hline
		Angle    & $\alpha$        & $\beta$        & $\gamma$       \\
		     & 90º & 90º    & 120º \\ \hline
		\multicolumn{4}{c}{Atomic positions (fractional)}     \\ \hline
		Te (4f)    & $1/3$        & $2/3$        & $0.09018$       \\
		Fe (4e)    & $0$        & $0$        & $0.1718$       \\
		Fe (2c)    & $1/3$        & $2/3$        & $1/4$       \\
		Ge (2d)    & $1/3$        & $2/3$        & $3/4$       \\
	\end{tabular}
\end{center}
\caption{Structural parameters of \FGT obtained from Ref.~\cite{structure}.}
\end{table}

\subsection{Bands without SOC}\label{ssec:sym_nosoc}

\FGT displays a ferromagnetic (FM) structure where the magnetic moments at Fe sites align parallel to the $c$ axis. The space group symmetry corresponds to SG $P6_3/mmc$ (No.194 in the Belov-Neronova-Smirnova notation\cite{bns_notation}). The Fe atoms sit at two different Wyckoff positions (WPs) labeled 2c and 4e. The calculations show that these two inequivalent positions host a magnetic moment per atom of approximately $1.5\,\mu_B$ and $2\,\mu_B$, respectively, which is in good agreement with the reported experimental measurement of $1.4\mu_B$ per atom.\\

In the absence of SOC (Fig.\ref{fig:bands_all}a), the Hamiltonian is decoupled into spin-up and spin-down blocks. As a consequence, each spin sector displays the symmetry of the single-valued gray group No.194.264\cite{spin-sectors}. This can be further justified by taking into account that the symmetry with FM order \textit{without SOC} is given by a Type-1 collinear SSG (L194.1.1) \cite{zhida_ssg}. Since in an SSG spatial and spin rotations are independent, the continuous rotation of the spin about the alignment axis decouples the bands into the two spin sectors. Each one obeys the symmetry of a discrete subgroup of the SSG which takes the form:
\begin{equation}
    G\times S_{Z^T_2},
\end{equation}
where $G$ contains only spatial operations in SG No.194 with identity spin part and $S_{Z^T_2}$ is generated by $\Theta U_x(\pi)$. Here, $\Theta$ denotes time-reversal while $U_x(\pi)$ is a $\pi$ rotation about the $x$ axis for the spins. $\Theta U_x(\pi)$ is represented by $i\sigma_zK$ (where $K$ denotes complex conjugation) and commutes with $G$. The operator acts as an analogue of TRS but, since $(\Theta U_x(\pi))^2 = +1$, each sector obeys the symmetry of a \textit{single-valued} gray group. Therefore, each spin sector can be analyzed separately as spinless fermions of a non-magnetic crystal.\\

The previously reported nodal line occurs along the high-symmetry line $P$ joining the points $K$ and $H$. Since along the nodal line two bands become degenerate, we must only consider 2D irreps at the endpoints, therefore leaving as possible coreps $K_5,K_6$ and $H_1,H_2,H_3$ respectively. Because the subduction of the coreps onto the $P$ line must match at both ends of the path, we can discard the $H_3$ symmetry. This is due to to $K_5$, $K_6$, $H_1$ and $H_2$ all subducing to $P_3$ whereas $H_3$ subduces to $P_1 \oplus P_2$, violating the compatibility relations. We also note that there are non-degenerate energies for a given spin-sector at $K$ (e.g. spin-down bands at around $-0.6$ eV) since there are one-dimensional coreps of the little group at that point. On the contrary, there are only 2D coreps at $H$ and all the bands are doubly degenerate.\\

The line $P=(1/3,1/3,w)$ is related to $(1/3,1/3,-w)$ by symmetry, which implies that the aforementioned twofold degeneracy forms a nodal line that closes due to periodic boundary conditions of the Brillouin zone (BZ). There are more symmetry mechanisms that protect nodal lines. In particular, as both spin sectors are decoupled, the crossing between up and down bands is never gapped and such accidental degeneracies can give rise to nodal lines\cite{nodal_lines}. Mirror symmetries can also protect band crossings inside each spin sector. The Bloch states on mirror-invariant planes in reciprocal space can be labeled according to the coreps of the little group of the plane. If this little group has more than one corep, the crossings of bands with different mirror eigenvalue are protected on the whole plane, which can give rise to nodal lines, as Hamiltonian matrix elements between states of different symmetry must be exactly zero, as we will next show. Given states $\ket{\psi(\kvec)}$ and $\ket{\phi(\kvec)}$ transforming as different coreps $D$ and $\tilde D$ of the little group, the action of the mirror plane $m$ is represented by unitary matrices $D(m)$ and $\tilde D(m)$ respectively. The matrix element of the Bloch Hamiltonian $H(\kvec)$ at points $\kvec_p$ on this plane $\bra{\psi(\kvec_p)}H(\kvec_p)\ket{\phi(\kvec_p)}$ must be invariant under this symmetry operation
\begin{equation}
    \bra{\psi(\kvec_p)}m^{-1}H(\kvec_p)m\ket{\phi(\kvec_p)} = \bra{\psi(\kvec_p)}H(\kvec_p)\ket{\phi(\kvec_p)},
\end{equation}
which follows because $m$ leaves points $\kvec_p$ on the plane invariant. If instead of transforming $H(\kvec_p)$, one equivalently applies the operator to the states, one obtains
\begin{equation}
\begin{split}
    \bra{\psi(\kvec_p)}m^{-1}H(\kvec_p)m\ket{\phi(\kvec_p)} = \\D(m)^{-1}\tilde D(m) \bra{\psi(\kvec_p)}H(\kvec)\ket{\phi(\kvec_p)},
\end{split}
\end{equation}
which only holds if $D$ and $\tilde D$ are equivalent coreps. It follows that states transforming under different coreps cannot hybridize and give rise to gaps. For example, the mirror symmetry $\{m_{001}|0,0,1/2\}$ leaves all $\kvec=(k_x,k_y,k_z=0,1/2)$ invariant. The little co-group is isomorphic to the point group 2'/m (No. 5.3.14), which is generated by $\{m_{001}|0,0,1/2\}$ and $\Theta\{\bar 1|\mathbf 0\}$. This point group has two different coreps $A'$ and $A''$ and thus the states on that $\kvec$ plane with different symmetry can cross and result in a connected line of crossing points. The rest of the possibilities are shown in Table \ref{tab:invariant_planes_nosoc}.

\begin{table}[]
\begin{tabular}{l|l|l|l}
Generators                                                      & Invariant plane       & Point group & Coreps   \\ \hline
$\{m_{001}|0,0,\frac{1}{2}\},\Theta\{\bar 1|\mathbf 0\}$        & $(k_x,k_y,k_z=0,1/2)$ &         &  \\
$\{m_{110}|\mathbf 0\},\Theta\{\bar 1|\mathbf 0\}$               & $(-k,k,k_z)$         &         &  \\
$\{m_{100}|\mathbf 0\},\Theta\{\bar 1|\mathbf 0\}$              & $(k_x=0,1/2,k_y,k_z)$ & 2'/m        & $A',A''$ \\
$\{m_{1\bar 1 0}|0,0,\frac{1}{2}\},\Theta\{\bar 1|\mathbf 0\}$ & $(k,k,k_z)$           &         &  \\
$\{m_{120}|0,0,\frac{1}{2}\},\Theta\{\bar 1|\mathbf 0\}$        & $(-2k,k,k_z)$         &         &  \\
$\{m_{210}|0,0,\frac{1}{2}\},\Theta\{\bar 1|\mathbf 0\}$        & $(-k,2k,k_z)$         &         & 
\end{tabular}
\caption{Generators of the little groups (with translations), invariant coordinates, little co-group label and coreps for the band structure without SOC that protect crossings that may lead to a nodal line. All invariant planes share the same point group and coreps.}
\label{tab:invariant_planes_nosoc}
\end{table}

\subsection{Bands with SOC}\label{ssec:sym_soc}

\begin{figure*}
    \centering
    \includegraphics{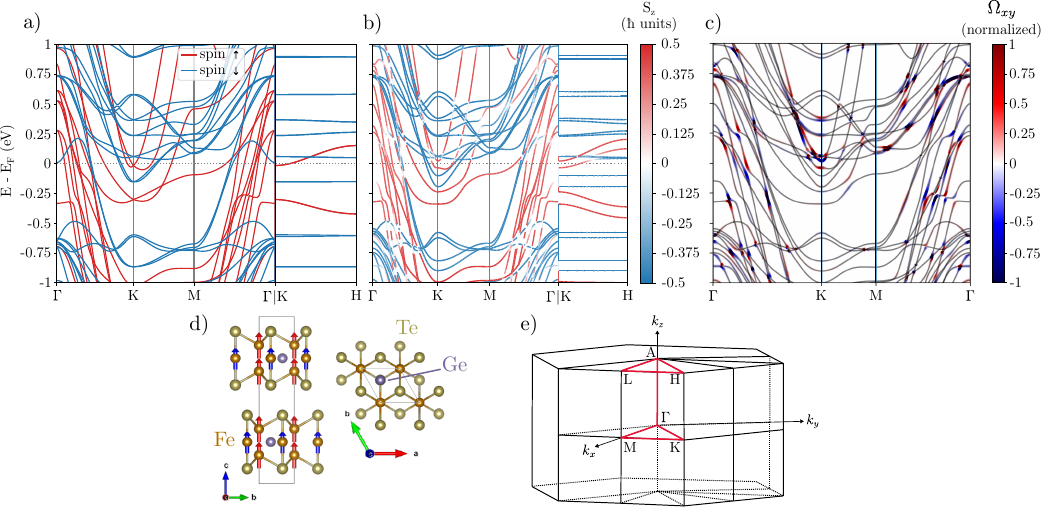}
    \caption{Electronic band structure along high-symmetry paths \textbf{a)} without SOC and colored spin sectors \textbf{b)} with SOC and $S_z$ spin component \textbf{c)} with SOC and normalized $\Omega_{xy}$ component of the Berry curvature \textbf{d)} Crystal structure as seen from the $a$ and $c$ directions. The arrows on the atoms represent the magnetic moments \textbf{e)} First BZ for the given structure with the high-symmetry path of the band plots marked in red.}
    \label{fig:bands_all}
\end{figure*}

When SOC is considered, spin and spatial operations cannot be decoupled. As a consequence, the symmetry group is reduced from the SSG to the Type-III MSG P6$_3$/mm'c' (No.194.270), which admits a coset decomposition
\begin{equation}
    M = G + \Theta\{2_{110}|\mathbf 0\}G,
\end{equation}
where $G$ is the unitary subgroup $P6_3/m$ (No.176) generated by $\{6^+_{001}|0,0,1/2\}$, $\{2_{001}|0,0,1/2\}$, $\{\bar 1|\mathbf 0\}$ and translations $\{1|\mathbf a_i\}$ by the vectors $\mathbf a_i$ defining the standard hexagonal unit cell. This MSG shares the Bravais lattice and symmetries of the unitary subgroup $P6_3/m$ (No.176), while also respecting additional space-group operations of No.194 combined with the antiunitary TRS. As shown in Figure \ref{fig:bands_all}b, the spin sectors remain well differentiated except at crossing points or points where a gap between spin up and down bands is opened, since the effect of SOC is relatively small. The states at each $\kvec$ point belong to a co-representation of the corresponding \textit{double} little group. Consequently, all previous degeneracies at $K$, $P$ and $H$ are lifted, since only double, one-dimensional coreps remain, and the nodal line is gapped.\\

The little group at points $\kvec = (k_x, k_y, 0\text{ or }1/2)$ is generated only by translations and the mirror plane $\{m_{001}|0,0,1/2\}$ and has two double-valued coreps with mirror eigenvalues $\pm i$. This means that, on either of the two planes, states with different mirror eigenvalue will not hybridize and can give raise to symmetry-protected nodal lines.

\section{Nodal lines}\label{sec:nodal_lines}
\subsection{Abundance of nodal crossings}\label{ssec:nodal_abundance}

\begin{figure*}
    \centering
    \includegraphics[width=0.95\linewidth]{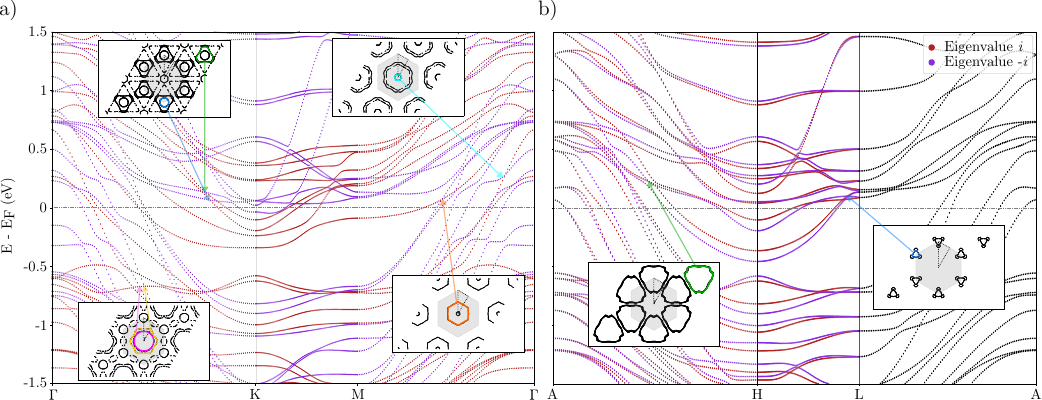}
    \caption{Nodal lines on the \textbf{a)} $k_z=0$ and \textbf{b)} $k_z=1/2$ planes. The bands are colored according to the mirror eigenvalue $i$ (red) and $-i$ (purple). The black dots indicate that along $L-A$ the states merge into a twofold-degenerate band with opposite mirror eigenvalues. The insets represent some of the detected nodal lines around the Fermi level. For each nodal line, an intersection with the high-symmetry path is marked by an arrow of the same color.}
    \label{fig:nodal_both}
\end{figure*}

To reveal the existence of nodal lines, we first calculate the crossing points between every pair of  consecutive bands in the TB model around the Fermi energy. The computation is restricted to the $k_z=0,1/2$ planes, where mirror symmetry protects the crossing of bands belonging to one of the two possible coreps, as explained in Subsec.\ref{ssec:sym_soc}. A selection of nodal lines is shown in Figure \ref{fig:nodal_both}, for $k_z=0,1/2$. On the $k_z=1/2$ plane, the mirror $\pm i$ states merge into the single doubly-degenerate irrep $\bar R_3 \bar R_4$ along the $A-L$ path. Consequently, the crossing points on that high-symmetry line cannot be part of a closed loop and, therefore, we find far fewer nodal lines on $k_z=1/2$ (for example, the nodal lines in Fig.\ref{fig:nodal_both}b do not intersect the $A-L$ path).

Symmetry-protected nodal lines can be characterized by computing the evolution along $k_y$ paths of Wilson loops computed along $k_z$\cite{drumhead_states}. Due to the mirror symmetry, the trace of the Wilson loop operator is quantized to either $0$ or $\pi$. When the lines are indeed protected, the trace experiments a $\pi$ jump, which is identified as a symmetry-protected nodal line that cannot be removed by a small, local perturbation that preserves the symmetry. This is because the value of the trace is related to the number of states with positive mirror eigenvalue\cite{drumhead_states}. However, we note that the actual value of the Wilson loop depends on the position of the mirror plane inside the unit cell, which is convention-dependent. For example, a shift of half the $c$ axis of all atomic sites would exchange the 0 and $\pi$ (modulo $2\pi$) values of Wilson loops of this kind. This can be understood by considering that, fixing $k_x$ and $k_y$, the remaining $k_z$ momentum corresponds to a 1D system analogous to a Su-Schrieffer-Heeger (SSH) model\cite{ssh}, where the value of the quantized Wilson loop of chains with and without dangling states can be exchanged by an origin shift in the unit cell. Alternatively, the trace of the Wilson line operator is proportional to the average $z$ position of the centers of the hybrid Wannier functions, extended in $x,y$ and localized the $z$ direction. From this point of view, it is clear that the choice of origin of the unit cell has such effect. The displacement of the atoms inside the unit cell can be regarded as a gauge transformation of the Bloch wave functions. Note that the argument above does not contradict the gauge invariance of the trace of a Wilson loop, since it applies for a set of bands that is completely isolated from the rest. This assumption falls apart in this case because we only track the lower band involved in every nodal crossing, which implies that the set of bands is not isolated.

\subsection{Drum-head states} \label{ssec:drumhead}

\begin{figure*}
    \centering
    \includegraphics{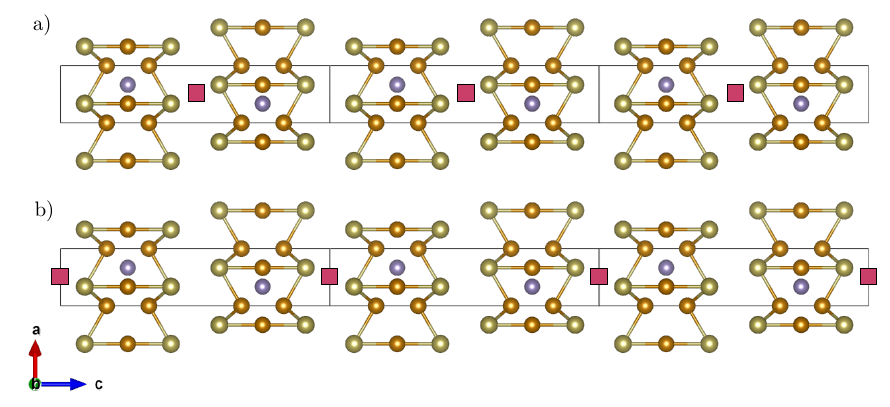}
    \caption{Sketch of the two possible positions of the hybrid Wannier function centers of the nodal lines for a slab of three cells in the $c$-axis direction. The colored spheres represent Fe (orange-brown) Te (brass) and Ge (purple) atoms. The red square is the position of the hybrid Wannier function of the occupied bands of a Wilson loop calculation along $k_z$. \textbf{a)} The trace of the Wilson loop is $\phi = 0$ and the HWF center is at $z=0$. \textbf{b)} $\phi = \pi$ and the center is at $z=1/2$, which gives rise to surface states since the surface termination is exactly at the HWF center.}
    \label{fig:drumhead_sketch}
\end{figure*}

\begin{figure*}
    \centering
    \includegraphics{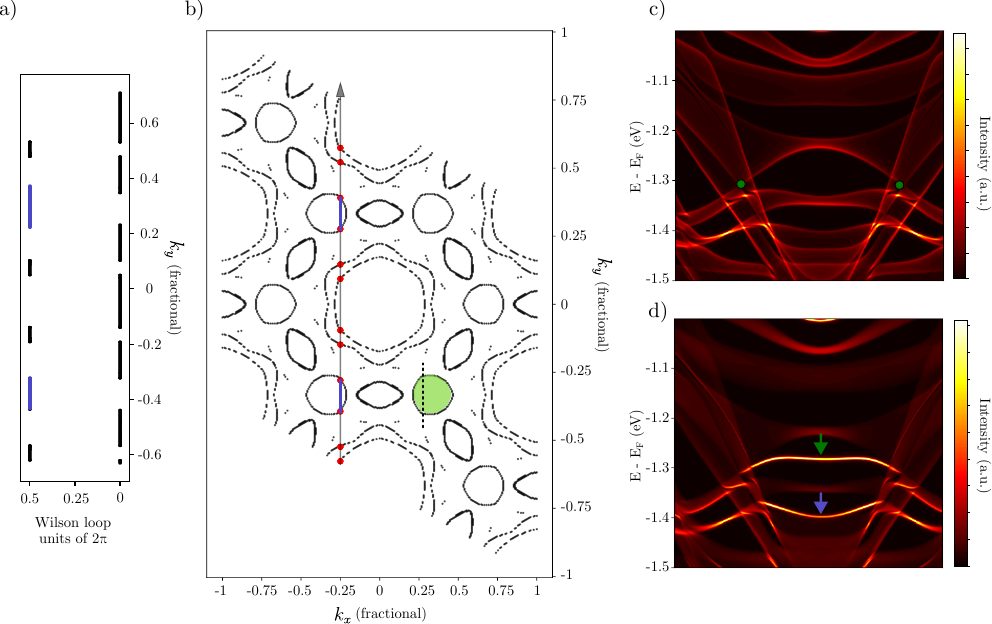}
    \caption{Characterization of one of the nodal lines and its drum-head state. \textbf{a)} Trace of Wilson loops along $k_z$ on a $k_y$-directed path indicated by an arrow in \textbf{b}. \textbf{b)} Gapless points between one pair of bands of the TB model on the $k_z=0$ plane. The green area marks the projection where the drum-head state resides. The red points mark the discontinuities in the Wilson loop evolution, which are a signal of a topologically protected nodal crossing. The blue segments help identify the points where the trace is $\phi=\pi$ (matched in \textbf{a}) inside this particular nodal line. \textbf{c)} Bulk spectrum for a semi-infinite slab cut perpendicular to the $c$ axis. The green dots are the intersections of the path in broken line in \textbf{b} with the nodal line. \textbf{d)} Surface spectrum where the corresponding drum-head state is indicated by a green arrow. Additionally, another drumhead state from a different nodal line is marked with a blue arrow (analysis not shown).}
    \label{fig:drumhead_state}
\end{figure*}

Nodal lines host drum-head states, which can be observed in the surface spectrum inside the surface projection of the nodal line provided that they are not hidden in the projection of the bulk states. Using the TB model, we computed the surface spectrum of semi-infinite slabs perpendicular to the $c$ axis using an iterative Green's function method\cite{surface_green_function}, as implemented in WannierTools. We were able to identify  at least one drum-head state can be observed extending over the area surrounding the $\Gamma$ point on the $k_z = 0$ plane, as show in Figure \ref{fig:drumhead_state}. The presence of surface states inside the projection of a nodal line can be predicted by the trace of the Wilson loop. Note however, that which value, either 0 or $\pi$, indicates the presence of drum-head states depends on whether the surface termination cuts the center of the corresponding $z$-localized hybrid Wannier functions (HWF). In our particular setting, a trace of $\phi = \pi$ serves as a confirmation that the observed state a drum-head state as sketched in Figure \ref{fig:drumhead_sketch}.

\section{Anomalous Hall Effect} \label{sec:ahc}

To analyze the AHC in \FGT, we compute it from the TB model in a range for 2 eV around the calculated Fermi energy using WannierBerri. Here, we identify three separate contributions that give rise to the final response in the material.

\subsection{Broken nodal line contribution}\label{ssec:broken_nodal}
Protected nodal lines like the ones shown in the previous section do not contribute to the AHC as they are not a source of Berry curvature\cite{kh_nodal_line}. However, the transition to the ferromagnetic order can gap nodal lines which are protected in the paramagnetic phase. Without magnetism, the symmetry of the system corresponds to the gray group No.194.264, which has mirror planes $\{m_{100}|\mathbf{0}\}$ and $\{m_{010}|\mathbf{0}\}$. These symmetries also protect nodal lines in $k_x=0,1/2$ and $k_y=0,1/2$ respectively but, once broken by the preferred magnetization axis, a small lifting of the nodal degeneracy occurs, which usually gives rise to a concentration of the Berry curvature at those points. The curvature due to this mechanism can contribute to the AHC signal in the real system.\\

\begin{figure}
    \centering
    \includegraphics{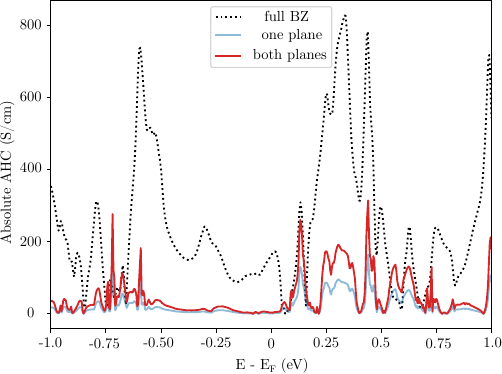}
    \caption{Contribution to the AHC by the nodal lines on $k_x=0,1/2$ (red) and $k_y=0,1/2$ (green). The signal from both sets of planes is identical since they are related by the symmetry of the gray group. The contribution from both sets (purple) does not account for the total response (doted line), especially in the range between -0.5 and 0.1 eV, which show that this is not the only source of AHC.}
    \label{fig:ahc_planes}
\end{figure}

To compute the contribution of these lifted nodal crossings, we perform AHC calculations in narrow momentum slices enclosing the mirror-invariant $k_x$ and $k_y$ planes. Figure \ref{fig:ahc_planes} shows the total AHC $\sigma_{xy}$ as a function of the chemical potential along with the contribution from the planes exclusively. It clearly shows that there must be other sources that give rise to the intrinsic AHC. One can also check using this method that the mirror-invariant $k_z=0,1/2$ planes, where the nodal lines are preserved, show no contribution . The plot also suggests that slightly shifting the chemical potential by 0.3 eV, for example by electron doping, could lead to a fourfold enhancement of the intrinsic AHC from around 200 S/cm to approximately 800 S/cm.

\subsection{Weyl nodes} \label{ssec:weyl_points}

\begin{figure*}
    \centering
    \includegraphics{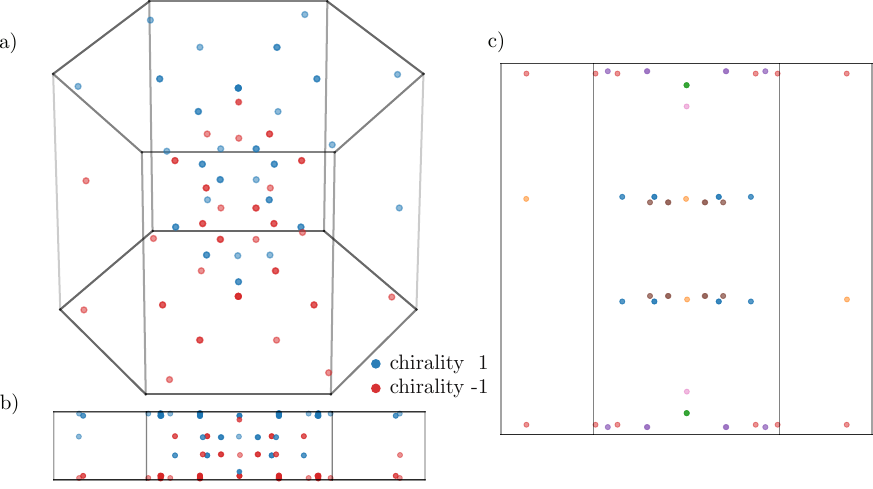}
    \caption{Weyl points within $\pm0.1$ eV from the Fermi level. \textbf{a)} Colored by charge $+1$ or $-1$ (red and blue respectively. The BZ is re-scaled in the $k_z$ direction for visibility purposes. \textbf{b)} Projection onto the $(k_y, k_z)$ plane with correct $k_z$ scale. \textbf{c)} $(k_y,k_z)$ projection of a colored by symmetry-related orbits.}
    \label{fig:weyl_points}
\end{figure*}

The AHC hosted by the system could also be explained by a large intrinsic contribution due to Weyl nodes in the band structure\cite{weyl_transport}. Hence, we search for Weyl crossings by finding gapless points between consecutive bands throughout the BZ and away from the mirror-protected $k_z=0,1/2$ planes. In contrast to nodal lines, which require symmetry protection, Weyl points will generically occur in three-dimensional crystals\cite{vanderbilt_book}. Their chirality is computed by a Wilson loop calculation over a sphere enclosing the point which is equivalent to computing the curvature flux through the sphere. Due to the limited resolution of the $\kvec$-grid used in the search process, it is not guaranteed to find all the Weyl points required by symmetry constraints. To circumvent this, we obtained the symmetrized points by applying all the symmetry operations of the MSG. The chirality of Weyl nodes related by symmetry obeys the following constraint regarding the proper/improper nature of the operation and its unitarity\cite{vanderbilt_book}:
\begin{align}
    \chi(\kvec) &\xrightarrow{\phantom{a}\text{inversion}\phantom{aa}} -\chi(-\kvec),\\
    \chi(\kvec) &\xrightarrow{\text{time-reversal}} \phantom{-}\chi(-\kvec),
\end{align}
where $\chi(\kvec)$ is the chirality of a node at point $\kvec$. This also allows us to group Weyl points related by symmetry into orbits whose number of nodes depends on the index of the little co-group of the wave vector in the point group of the crystal. Figure \ref{fig:weyl_points} shows all the Weyl points within a 0.1 eV range from the Fermi level, grouped by orbits. One can also check that, after the symmetrization process, the total chirality of all the points is zero, as required by the Nielsen-Ninomiya theorem\cite{nielsen_ninomiya}.

\begin{figure}
    \centering
    \includegraphics{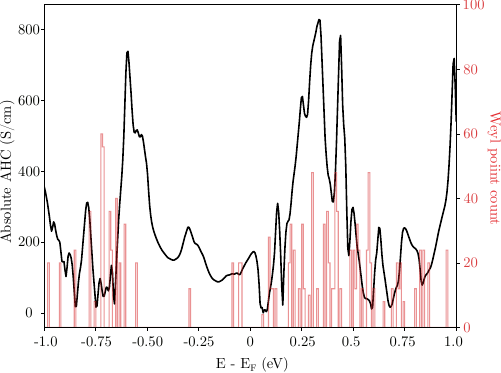}
    \caption{Computed absolute AHC $\sigma_{xy}$ (black) and the energy distribution of the non-zero-chirality Weyl nodes (red) in a window extending 1.5 eV above and below the Fermi energy.}
    \label{fig:ahc_weyl}
\end{figure}

To relate the intrinsic AHC to the numerous Weyl points in the band structure, we plot in Figure \ref{fig:ahc_weyl} the AHC along with the energy distribution of the Weyl nodes in a window 1 eV above and below the Fermi level. We observe that the change in AHC above the Fermi level is correlated to the presence of Weyl nodes in that energy range. However, the plot also shows that nodal points cannot explain the features in the interval around -0.6 eV to -0.1 eV, where they are almost absent. As one moves the chemical potential, the Weyl crossings that are populated contribute to the final response and, as such, their presence at a certain energy should correlate with a change in the AHC. We emphasize that their effect can be either to suppress or enhance the AHC, depending on details of the actual electronic band structure.

\subsection{Gaps due to SOC} \label{ssec:soc_gaps}
\begin{figure}
    \centering
    \includegraphics[width=0.95\linewidth]{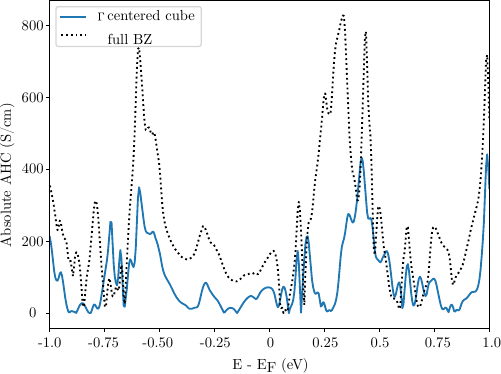}
    \caption{AHC from a cube enclosing the $\Gamma$ point (blue), which estimates the contribution of the SOC-related gaps as seen in Figure \ref{fig:bands_all}c. It shows that the signal in the -0.5 to 0.1 eV range around the Fermi level can be attributed to this mechanism.}
    \label{fig:ahc_cube}
\end{figure}

As shown in Section \ref{sec:symmetry}, SOC mixes the spin components of the states and can lift the degeneracy in some cases, such as the specific nodal line on the $K-H$ path, even when the degenerate states are purely spin-up or down\cite{double_space_group}. In addition to this source, one must also consider that, without SOC, the spin sectors are completely decoupled and bands of different spin are free to cross. Once SOC is taken into account, both spin orientations are coupled, generically giving rise to gaps\cite{soc_gap} where the Berry curvature can concentrate and thus contribute to the AHC (compare Fig.\ref{fig:bands_all}a and Fig.\ref{fig:bands_all}b). This can be proven by computing the Berry curvature $\Omega$ on high-symmetry paths, as shown in Figure \ref{fig:bands_all}c, defined by
\begin{equation}
    \Omega_{\alpha\beta} = \partial_\alpha A_\beta - \partial_\beta A_\alpha - i[A_\alpha,A_\beta],
\end{equation}
where $A$ is the Berry connection, $\partial_\alpha$ is the partial derivative with respect to $k_\alpha$ and $[\cdot\,,\cdot]$ is the commutator, and then obtaining the AHC as
\begin{equation}
    \sigma_{\alpha\beta} = -\frac{e^2}{\hbar}\int \frac{d^3k}{(2\pi)^3}\Omega_{\alpha\beta}(\kvec).
\end{equation}
The plot shows that many of these avoided crossings are located around $\Gamma$, which allows us to estimate their contribution to the AHC by calculating the anomalous conductivity in a cube surrounding this point, as show in Figure \ref{fig:ahc_cube}. This reveals that SOC-induced gaps are the main mechanism giving rise to the signal observed in the range from -0.6 eV to -0.1 eV approximately, along with the already computed contribution from the lifted nodal degeneracies.

\section{Conclusion} \label{sec:conclusion}

In this work, we have analyzed the symmetry of \FGT in-depth and performed DFT and Wannier-based modeling calculations. Full understanding of the symmetry with and without SOC and for the paramagnetic and FM phases of the material allows the prediction of topological features and to explain the observed anomalous Hall conductivity. The symmetry of the FM phase, with SOC considered, is described by the MSG No.194.270 whose $\{m_{001}|0,0,1/2\}$ mirror plane, in particular, protects nodal lines on the invariant $k_x=0,1/2$ planes. We have found that \FGT hosts many of these features (Fig.\ref{fig:nodal_both}) and given a rigorous identification of the nodal crossings through the discontinuities of Wilson loops over paths that intersect them. We have used the trace of the Wilson loop operator to predict the appearance of drum-head surface states in the projection of the nodal lines and have detected one example in the computed surface DOS (Fig.\ref{fig:drumhead_state}).
Furthermore, we have computed the AHC for a range of chemical potentials and detected the three main mechanisms that give rise to the AHC response. Firstly, while the paramagnetic phase also protects nodal lines on $k_x=0,1/2$ and $k_y=0,1/2$, the FM transition gaps them and they contribute to the anomalous transport due to being points where the Berry curvature concentrates (Fig.\ref{fig:ahc_planes}). However, they are not the only source of AHC, since its contribution is far from the total response computed. Secondly, we have identified many Weyl points in the band structure and computed their chirality (Fig.\ref{fig:weyl_points}). Their contribution to the AHC has to be complemented by the two other sources, especially in the -0.6 to 0.1 eV range, where Weyl crossings are almost absent (Fig.\ref{fig:ahc_weyl}). Finally, with magnetic order but neglected SOC, the spin up and down bands are decoupled and free to cross, each set respecting the symmetries of the single-valued gray MSG No.194.264. These degeneracies are lifted when SOC is brought into the picture, giving rise to small gaps where the curvature is also prominent (Fig.\ref{fig:bands_all}c and \ref{fig:ahc_cube}). The calculations also show that slightly shifting the chemical potential by approximately 0.3 eV would further increase the AHC in \FGT from around 200 S/cm to approximately 800 S/cm, suggesting electron doping as a good mechanism to obtain an enhanced effect.

\section{Acknowledgements} 
M.G.D., I.R. and M.G.V. thanks support to the Spanish Ministerio de Ciencia e Innovación (PID2022-142008NB-I00), the Canada Excellence Research Chairs Program for Topological Quantum Matter and funding from the IKUR Strategy under the collaboration agreement between Ikerbasque Foundation and DIPC on behalf of the Department of Education of the Basque Government. M.G.D. acknowledges financial support from Government of the Basque Country through the pre-doctoral fellowship PRE 2023 2 0024.

\bibliography{bibliography}
\end{document}